\documentclass[aps, twocolumn,prx,showkeys,showpacs, longbibliography, citeautoscript]{revtex4-2}

\usepackage{graphpap}
\usepackage[dvips]{graphicx}
\usepackage[dvips]{graphics}
\usepackage{xcolor}
\usepackage{textcomp}
\usepackage{amsmath, amssymb, amsfonts, amsthm}
\usepackage{braket}
\usepackage{bbold}
\usepackage[pagebackref=false]{hyperref}
\usepackage{xr}
\usepackage{CJK} 
\usepackage{soul}
\usepackage[pagewise]{lineno}
\makeatletter
\newcommand*{\addFileDependency}[1]{
  \typeout{(#1)}
  \@addtofilelist{#1}
  \IfFileExists{#1}{}{\typeout{No file #1.}}
}
\makeatother
\RequirePackage[T1]{fontenc}
\RequirePackage[utf8]{inputenc}
\RequirePackage{lmodern}

\begin{document}

\title{Shuttling an electron spin through a silicon quantum dot array}

\author{A.M.J.~Zwerver$^1$}
\author{S.V.~Amitonov$^2$}
\author{S.L.~de Snoo$^1$}
\author{M.T.~M\k{a}dzik$^1$}
\author{M.~Russ$^1$}
\author{A.~Sammak$^{2}$}
\author{G.~Scappucci$^{1}$}
\author{L.M.K.~Vandersypen$^{1*}$}
\affiliation{$^1$QuTech and Kavli Institute of Nanoscience, Delft University of Technology, Lorentzweg 1, 2628 CJ Delft, The Netherlands\\
$^2$ QuTech and Netherlands Organization for Applied Scientific Research (TNO), Delft, The Netherlands}

\date{\today}
\begin{abstract}
\begin{center}
\emph{$^*$ Corresponding author: l.m.k.vandersypen@tudelft.nl}
\end{center}

Coherent links between qubits separated by tens of micrometers are expected to facilitate scalable quantum computing architectures for spin qubits in electrically-defined quantum dots. These links create space for classical on-chip control electronics between qubit arrays, which can help to alleviate the so-called wiring bottleneck. A promising method of achieving coherent links between distant spin qubits consists of shuttling the spin through an array of quantum dots.
Here, we use a linear array of four tunnel-coupled quantum dots in a $^{28}$Si/SiGe heterostructure to create a short quantum link. We move an electron spin through the quantum dot array by adjusting the electrochemical potential for each quantum dot sequentially. By pulsing the gates repeatedly, we shuttle an electron forward and backward through the array up to $250$ times, which corresponds to a total distance of approximately $80$ \textmu m. We make an estimate of the spin-flip probability per hop in these experiments and conclude that this is well below $0.01\%$ per hop.
\end{abstract}
\maketitle
\newpage

\section{I. Introduction}

Practical quantum computation requires millions of high-quality qubits~\cite{reihera_elucidating_2017}. In light of this, spin qubits in electrically-defined quantum dots are of particular interest because they combine a high quality with a small footprint (100 nm pitch)~\cite{vandersypen_interfacing_2017}. Recently, long-lived spin coherence~\cite{veldhorst_addressable_2014}, high-fidelity single-qubit gates~\cite{yoneda_quantum-dot_2018, yang_silicon_2019, lawrie_simultaneous_2021} and high-fidelity two-qubit gates~\cite{xue_quantum_2022, noiri_fast_2022, mills_two-qubit_2022} have all been demonstrated, as well as universal control of up to six qubits~\cite{hendrickx_four-qubit_2021,philips_universal_2022}. Moreover, a high fabrication yield can be achieved by employing semiconductor manufacturing techniques~\cite{maurand_cmos_2016,zwerver_qubits_2022}. 

It is commonly recognized that for spin qubits, long-range on-chip qubit connections are a crucial part of future devices containing millions of qubits~\cite{taylor_fault-tolerant_2005, vandersypen_interfacing_2017, li_crossbar_2018, buonacorsi_network_2019, boter_spider-web_2022}. The underlying reason is that in current implementations, at least one control wire is routed from off-chip electronics to every qubit on the chip, an approach that becomes impractical beyond thousands or tens of thousands of qubits~\cite{franke_rents_2019}. To scale beyond, the control signals can be delivered to the chip through a fixed number of wires and distributed to the qubits by classical on-chip electronics. On-chip quantum links would allow to create space between qubit registers, say on the 10 \textmu m needed for this classical electronics.

\begin{figure*}[t]
    \centering
    \includegraphics[width=\textwidth]{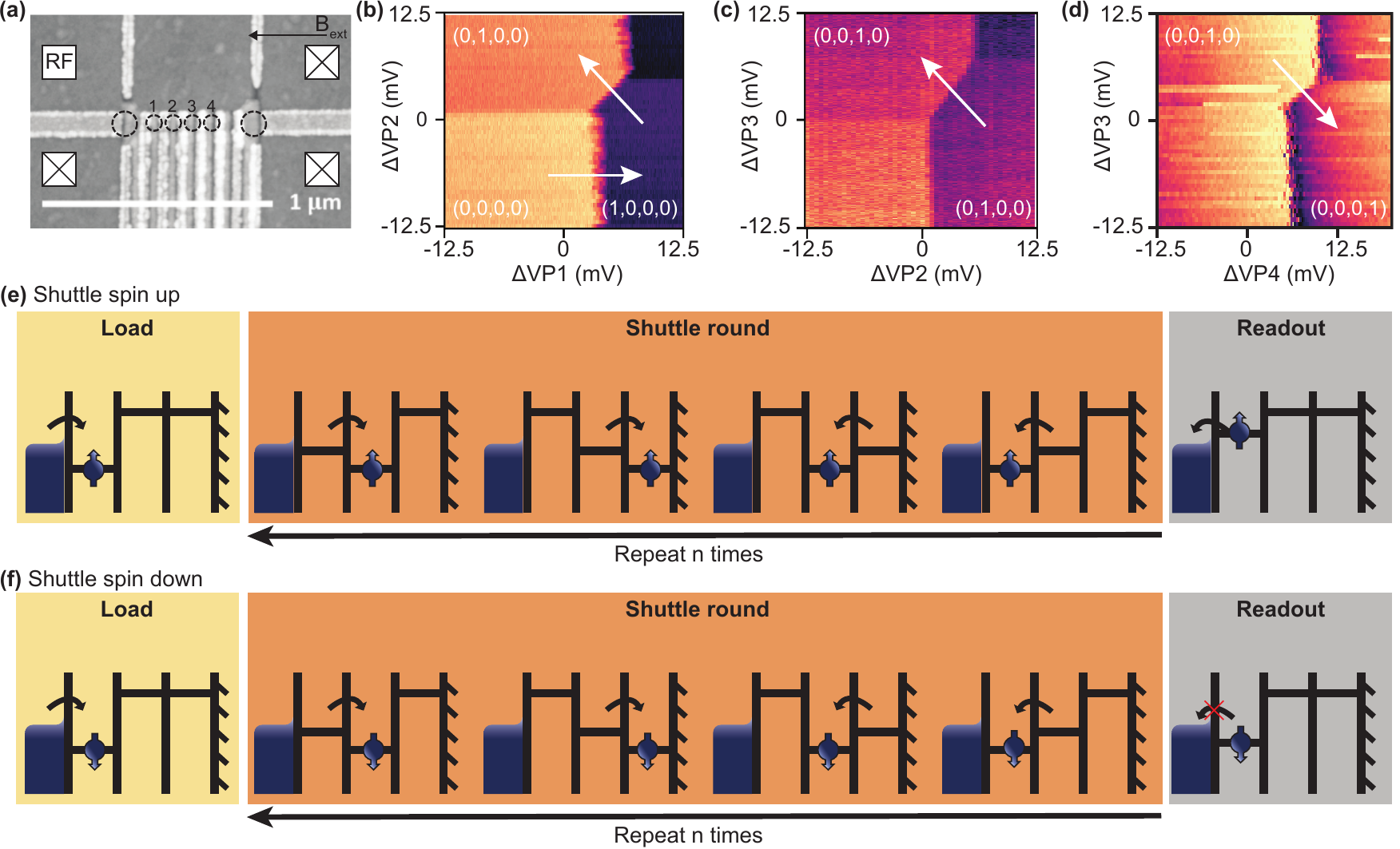}
    \caption{Shuttling through a Si/SiGe device. (a) Scanning-electron microscope image of a device nominally identical to the device used for this work. Plunger and barrier gates are interleaved and have been fabricated in separate steps. The quantum dots used in this work are indicated by the small circles. The sensing dots on either side of the array are indicated by big circles. The left sensor is connected to an RF readout circuit via one of the ohmic reservoirs, indicated by squares. This sample does not contain a micromagnet. (b), (c), (d) Charge stability diagrams for (b) quantum dots 1-2 while quantum dots 3, 4 are empty, (c) quantum dots 2-3 while quantum dots 1,4 are empty and (d) quantum dots 4-3 while quantum dots 1,2 are empty (right sensing dot response). The charge stability diagrams are measured by means of virtual gates. For reference, 25 mV in the virtual gate voltage $\Delta VP1$ corresponds to 3.75 meV in the  electrochemical potential of dot 1. The electron occupation per charge region and the direction of the shuttling events are indicated in the figures. (e), (f) Shuttling procedure through three quantum dots. An electron with random spin is injected during the loading stage. Panels (e) and (f) show the case of a spin up and spin down electron respectively. At the end of the shuttling sequence, the spin state is read out using energy-selective tunnelling, where a spin-up electron will tunnel out of the quantum dot (e), whereas for a spin-down electron this is energetically forbidden (f). The sensing dot signal reveals whether or not the electron tunneled to the reservoir, from which the spin state is inferred.}
    \label{Figure_1}
\end{figure*}

Several approaches for on-chip quantum links between spin qubit registers have been proposed and explored. In one approach, qubit information is transferred along a qubit chain via coherent SWAP operations~\cite{friesen_efficient_2007, kandel_coherent_2019}. Alternatively, spin qubits could be coherently coupled at a distance by means of microwave photons in superconducting resonators~\cite{Harvey_coupling_2018, mi_coherent_2018, samkharadze_strong_2018, landig_coherent_2018, borjans_resonant_2020, harvey-collard_circuit_2021}. The footprint of the superconducting resonator makes this approach particularly interesting for qubit coupling beyond the 100~\textmu m-scale. In a third approach, quantum information transfer is achieved by physically displacing the electrons while preserving the spin state. This has been realized in GaAs by means of surface acoustic waves along up to $20$-\textmu m-long channels~\cite{mcneil_-demand_2011, hermelin_electrons_2011, jadot_distant_2021, Edlbauer_in-flight_2021}. This method makes use of the piezo-electro effect in GaAs, which is absent in Si. 

Electron shuttling propagated by gate voltages offers a promising alternative for spin transfer in silicon on the sub-mm scale. The most resource-efficient approach is the so-called conveyor belt method, where just four phase-shifted sine wave signals must be applied to a large number of gates to generate a traveling wave potential along a channel between qubit registers~\cite{langrock_blueprint_2022}. This conveyor belt method requires a high degree of uniformity in the channel but has recently been demonstrated in slow-motion~\cite{seidler_conveyor-mode_2021}. Bucket-brigade-mode shuttling, where an electron is transported through an array of quantum dots by successively adjusting their electrochemical potentials, offers a more accessible means of exploring spin shuttling, since local potential variations from background disorder can be compensated by individual gate voltages. This method has been successfully demonstrated in GaAs devices, both in linear quantum dot arrays~\cite{baart_single-spin_2016, fujita_coherent_2017} and in triangular~\cite{flentje_coherent_2017} and two-dimensional arrays~\cite{mortemousque_enhanced_2021}. Yet, with the focus of the research field shifting to group IV platforms for their superior coherence times because of spin-free isotopes, it is important to demonstrate shuttling in silicon devices. Coherent spin shuttling in silicon poses several new challenges~\cite{krzywda_interplay_2021, langrock_blueprint_2022}.  Silicon samples inherently exhibit more electrostatic disorder than GaAs samples, which can hinder the shuttling success. Moreover, Si has an extra degree of freedom in the form of the valley~\cite{zwanenburg_silicon_2013}, which adds an additional loss channel for the moving spin~\cite{friesen_valley_2007}. Promising results have been obtained by shuttling electrons through a linear array of nine Si/SiGe quantum dots with the bucket brigade mode, but without probing the impact on the spin states~\cite{mills_shuttling_2019}. So far, coherent spin shuttling in silicon devices has been limited to double dot experiments~\cite{yoneda_coherent_2021,noiri_shuttling_2022} and spin shuttling through extended quantum dot arrays remains to be demonstrated.

Here, we shuttle a single electron spin through a five-quantum-dot linear array, occupying three or four quantum dots. We move the electron forth and back through the array for up to $1000$ hops, equal to a distance of $80$~\textmu m, and test both whether the charges are transferred as expected and whether the spin polarization is preserved. 

\begin{figure*}[t]
    \centering
    \includegraphics[width=0.99\textwidth]{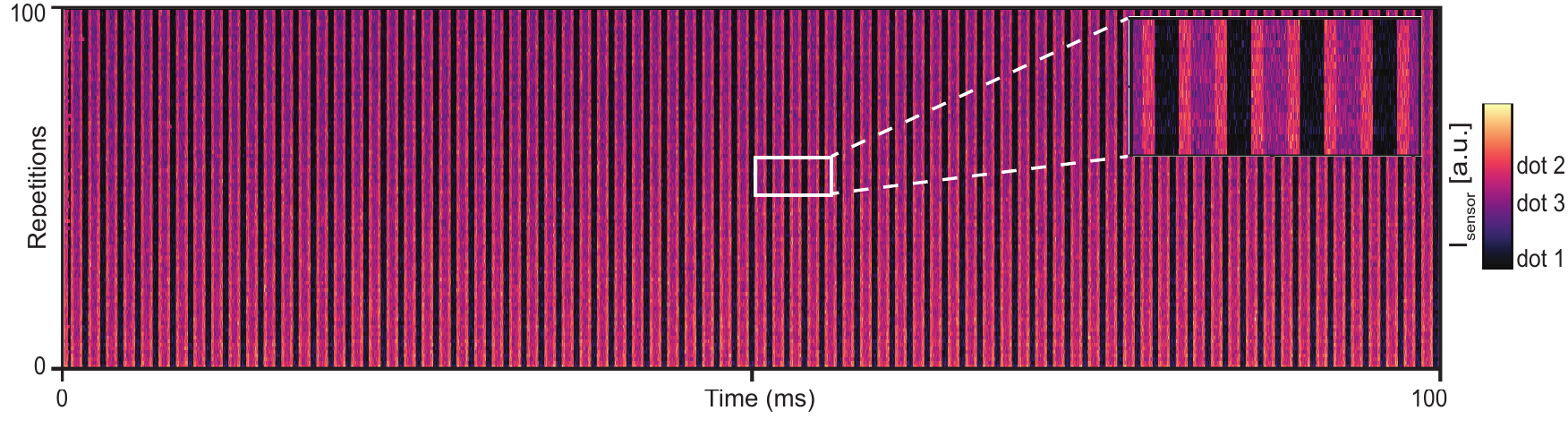}
    \caption{Charge shuttling over many rounds. Single-shot sensor response for $100$ shuttling traces. The spin is loaded in dot $1$ and then shuttled forth and back to dot $3$ for 78 rounds. Each single-shot trace has an offset with respect to the previous trace of $0.04$~mV in $VP_1$. Inset: zoom in of a part of the plot to better visualize the sensor response. We see that for each trace, the electron shuttles back and forth repeatedly as expected.}
    \label{fig: Figure_2}
\end{figure*}

\section{II. Device and shuttling procedure}

The quantum dot array (see Fig.~\ref{Figure_1}{a}) is fabricated on an undoped $^{28}$Si/SiGe heterostructure, with the quantum well positioned $\simeq 30$ nm below the surface~\cite{xue_cmos-based_2021, wuetz_atomic_2021}. The device contains three layers of Ti/Pd gate electrodes, used respectively as screening gates, plunger and accumulation gates, and barrier gates, isolated by Al$_2$O$_3$ dielectric layers in between. The device can host a linear array of up to five quantum dots with an $80$-nm pitch. The dots are formed by applying a bias to the metallic gate electrodes, where the plunger gates allow for adjustment of the electrochemical potentials of the quantum dots and the barrier gates control the tunnel couplings between the quantum dots. The screening gates define the lateral boundary of the quantum dot array. On either side of the array, a sensing dot is formed, which doubles as electron reservoir. Unless otherwise noted, all charge sensing data plots the response of the left sensing dot. All measurements are performed in a dilution refrigerator with a base temperature below $20$~mK and with an applied in-plane magnetic field $B_0$ of $1.3$~T. This magnetic field gives a Zeeman splitting of 150~\textmu eV, well above the valley splittings in the quantum dots, which are about $80$-$86$~\textmu eV (see Fig. S5a-c in Ref.~\cite{wuetz_atomic_2021}).

By means of virtual gates, we tune the five-quantum-dot linear array to the single-electron regime~\cite{mills_shuttling_2019, volk_loading_2019}. All quantum dots can be formed simultaneously, yet a defect in one of the high-frequency lines connected to the sample prohibits fast control of the fifth quantum dot. As these fast pulses are essential for shuttling, we will focus on the leftmost four quantum dots for the remainder of this work. We shuttle over both three quantum dots and four quantum dots (see Fig.~\ref{Figure_1}{b-d} for the charge stability diagrams). The tunnel coupling between the quantum dots is tuned in the range of $\sim3-7$~GHz (Appendix~\ref{Appendix G: tunnel coupling}).

A schematic of the shuttling sequence over three quantum dots is depicted in Fig.~\ref{Figure_1}{e,f}. The shuttling procedure starts in the (0, 0, 0)-regime, with the three electrochemical potentials of the quantum dots well above the Fermi energy of the reservoir. By pulsing the electrochemical potential of quantum dot 1 deep into the (1,0,0)-regime, we load an electron with random spin polarization in quantum dot 1. Then, we sequentially shuttle the electron through the 1-2 anticrossing into the (0,1,0)-regime and through the 2-3 anticrossing into (0,0,1). By reversing the sequence, we shuttle the spin back to quantum dot 1.

We repeat the shuttling sequence up to $n$ rounds, ending with the electron in dot 1, and then determine the spin polarization by means of energy-selective readout~\cite{elzerman_single-shot_2004}. The shuttling procedure through four quantum dots is similar, with the addition of a fourth quantum dot. To ensure a well-controlled transition between the quantum dots, we aim to move the electrons through the anticrossings adiabatically with respect to the inter-dot tunnel coupling. To this effect, we first abruptly pulse to a point close to but before the anticrossing, then ramp through the anticrossing (linear ramp of $2$~\textmu s over a $\approx 300$~\textmu eV detuning range, depending on the dot pair) before we abruptly pulse far beyond the anticrossing. The electrochemical potentials of the quantum dots that are meant to be unoccupied are pulsed well above the Fermi energy of the reservoir, to prevent the electron from tunnelling back and to prevent a second electron from entering the array. Shuttling the electron back and forth through multiple dots over multiple rounds, mimics shuttling through an extended quantum link and will be used to test how well the spin polarization is preserved.

\section{III. Spin shuttling}

We first verify whether shuttling the electron charge, disregarding the spin state, can be achieved reliably in this sample. To this end, we monitor the (virtual) sensor response while applying a sequence of pulses designed to shuttle the electron through the first three dots for 78 rounds (typical pulse amplitudes are $10$-$30$~mV in virtual gate voltage, and $10.5$-$35.1$~mV in real gate voltage). Apart from some latching during the initial electron loading in the array -- the tunnel barrier with the reservoir had to be sufficiently closed for the readout protocol to work well -- we see the expected sensor response, corresponding to the electron repeatedly occupying the three dots in the sequence 1-2-3-2-1  (Fig.~\ref{fig: Figure_2}). In this data set, the time spent in each dot in between hops is about $420$~\textmu s for dots 1 and 3, and half as long for dot 2 such that the total time spent in each dot is the same (Appendix~\ref{Appendix F: shuttle many rounds} shows similar data with about $100$~\textmu s between hops). The figure shows 100 repetitions of the charge shuttling experiment, each offset by 0.04 mV in $VP1$, showing some robustness of the protocol to gate voltage offsets. Reliable charge shuttling is a prerequisite for the electron spin to be shuttled with high fidelity.

\begin{figure*}[t]
    \centering
    \includegraphics[width=0.99\textwidth]{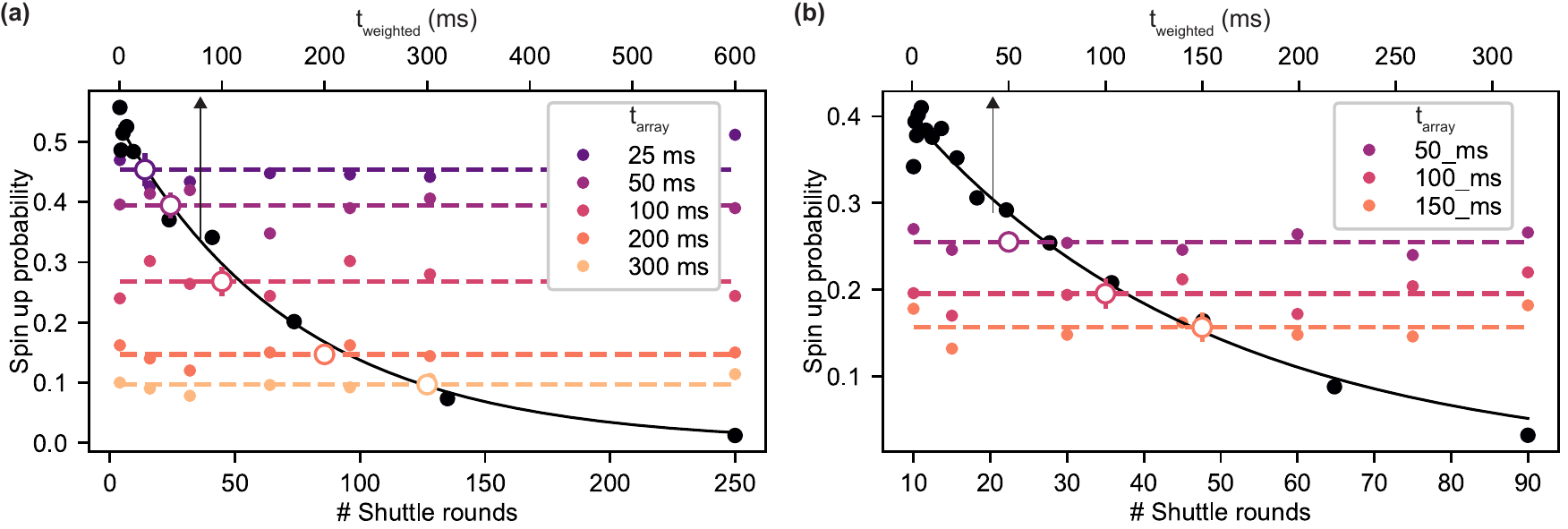}
    \caption{Shuttling through (a) three (b) four quantum dots. The coloured circles
    represent the spin-up probability for different times $t_{\mathrm{array}}$ after shuttling the spin back and forth through the array for $n$ rounds (bottom axis). Each data point is averaged 500 times. The dashed coloured lines represent the average of the data points for each $t_{\mathrm{array}}$ and are a guide to the eye. This average is also plotted as a function of total time spent in the array (top axis, open circles) with an error bar of one standard deviation (error bars sometimes fall behind the circles). The black circles show the spin-up probability after shuttling back and forth through the quantum dot array once as a function of the total time spent in the array (top axis). Each data point is averaged (a) 750, (b) 500 times. Fitting an exponential to the data yields a weighted $T_1$ of (a) $170\pm18$~ms (b) $156\pm33$~ms.}
    \label{fig: Figure_3}
\end{figure*}

Next, we aim to test to what extent the spin polarization is preserved during shuttling, following a similar procedure as used in earlier spin shuttling work in a GaAs device~\cite{baart_single-spin_2016}. Given that shuttling through the array over multiple rounds takes time, the spin state will gradually relax to the ground state. The question is whether additional spin flips, beyond those from spontaneous relaxation, are induced by the act of shuttling itself.

As a reference, we first measure the spin relaxation time in each quantum dot separately. We do so by following the shuttling procedure to move the spin to the desired quantum dot (in this case using $10$~\textmu s ramp times) and varying the wait time in this quantum dot. Thereafter, the electron is shuttled back to quantum dot 1 and its spin state is measured. Repeating this procedure multiple times, we can determine the average spin polarization. For the measurements in quantum dot 2 and 3, the load time in quantum dot 1 is set to $50$~\textmu s, to ensure both the loading of an electron in the quantum dot array and potential relaxation to the lower valley~\cite{yang_spin-valley_2013, watson_programmable_2018, borjans_single-spin_2019, cai_coherent_2021}. We determine a relaxation time for quantum dots 1, 2 and 3 of $T_{1,1}=129\pm33$~ms, $T_{1,2}=257\pm79$~ms and $T_{1,3}=152\pm48$~ms respectively, where the uncertainty is one standard deviation (see Appendix~\ref{Appendix E: relaxation times}).

Then, we shuttle the electron forth and back through the array once, while the total wait time in the array, $t_{\mathrm{array}}$, is divided symmetrically and equally over the quantum dots. By varying $t_{\mathrm{array}}$, the weighted relaxation time of the spin in the array can be determined. The weighted relaxation time for three dots is plotted in Fig.~\ref{fig: Figure_3}{a} (black circles). Fitting the data to an exponential yields a weighted relaxation time of $T_{1,\textrm{weighted}}=170\pm18$~ms. We expect the weighted $T_1$ to correspond to $3(1/T_{1,1}+1/T_{1,2}+1/T_{1,3})^{-1} = 164\pm47$~ms. The measured weighted relaxation time falls within the uncertainty range.

Finally, to analyse the direct effect of shuttling on the spin polarization, we vary both the number of shuttling rounds and the total shuttling time. We load an electron with random spin polarization into the array and shuttle it for a variable number of shuttle rounds through the array for a fixed total duration, $t_{\mathrm{array}}$. Figure~\ref{fig: Figure_3}{a (b)} shows the spin-up probability after $n$ shuttling rounds through three (four) dots for $t_{\mathrm{array}}$ varying between $25$~ms and $300$~ms. Note that each shuttle round through three (four) quantum dots contains four (six) subsequent hops between neighbouring quantum dots. For shuttling through three quantum dots, we find that, for each wait time, there is no sign of systematic decay of the spin-up probability as a function of the number of shuttle rounds up to the maximum of $n=250$ ($n=500$, see Appendix~\ref{Appendix D: shuttle 500 rounds}) rounds tested. This corresponds to $1000$ ($2000$) hops, or $80$ ($160$)~\textmu m of total distance traveled. For shuttling through four quantum dots, the results are analogous up to at least $90$ rounds ($540$ hops and $43$~\textmu m). In addition, the average spin up probability for all shuttling rounds per $t_{\mathrm{array}}$ (shown as the open circles) falls almost exactly on the weighted $T_1$ plot, which further points at the absence of spin flips, other than through spontaneous relaxation during wait times between hops. 

The apparent very low spin-flip probability from shuttling is consistent with the observations of Ref.~\cite{yoneda_coherent_2021} and is also expected considering the known mechanisms for spin flips. Electron exchange with the reservoirs should be suppressed because the dot-reservoir tunnel rate was set to a low value of 32~kHz. Spin flips from hyperfine interaction with nuclear spins should also be negligible in this quantum dot defined in an isotopically-enriched $^{28}$Si (800 ppm of $^{29}$Si) quantum well. The spin-orbit interaction is believed to be small in silicon samples, especially in Si/SiGe~\cite{zwanenburg_silicon_2013, prada_spin-orbit_2011}, although there are measurements in SiMOS indicating a spin-orbit length of only $1$~\textmu m~\cite{harvey-collard_spin-orbit_2019}.
Finally, we try to obtain a lower bound on the spin-flip probability per hop based on the data of Fig.~\ref{fig: Figure_3}. We analyse the accumulation of errors through transition-induced spin flips by simulating such events and comparing the simulations to the measured data. Even when the probability for transition-induced spin flips from both spin down to spin up and from spin up to spin down were as small as $0.01\%$ per transition, we should have seen an increase of spin-up probability with the number of shuttle rounds, see Appendix~\ref{Appendix C: simulations}.

The fastest wait times between shuttling used in this experiment amount to about $12.5$~\textmu s (500 rounds with four hops per round in 25 ms total). This time scale hovers around the $T_2^*$-times measured in isotopically enriched silicon~\cite{veldhorst_addressable_2014, yoneda_quantum-dot_2018, sigillito_coherent_2019, huang_fidelity_2019, struck_low-frequency_2020, xue_quantum_2022}. For coherent spin shuttling, the wait times should be reduced by at least one order of magnitude, and also the shuttling times ($2$~\textmu s ramp) should be brought down. 

\section{IV. Conclusion}
In conclusion, we demonstrated the shuttling of an electron through four quantum dots to mimic an array of up to $1000$ quantum dots in the bucket-brigade mode, with a spin-flip probability per hop well below 0.01\%. The method used in this work can be easily extended to larger quantum dot arrays, which highlights the potential of electron shuttling for long-range quantum links. Subsequent shuttle efforts should focus on decreasing the shuttle time by increasing the tunnel rate between the quantum dots and increasing the ramp rates through the anticrossings, which should all be feasible. In this way, the effect on the qubit coherence of shuttling through extended quantum dot arrays can be studied. In addition, it will be worthwhile to explore high-speed conveyor-belt shuttlers, propagating the electron in a travelling wave potential~\cite{seidler_conveyor-mode_2021, langrock_blueprint_2022}.

\section*{Acknowledgements}
We thank Lucas Peters and Zhongyi Jiang for useful input from measuring test structures, Raymond Schouten and Marijn Tiggelman for technical support and Inga Seidler and Lars Schreiber for useful discussions. Moreover, we thank everyone in the QuTech spin qubit group for input and support. We acknowledge financial support from the QuantERA ERA-NET Cofund in Quantum Technologies implemented within the European Union's Horizon 2020 Program and from Intel Corporation. This research was sponsored by the Army Research Office (ARO) under grant numbers W911NF-17-1-0274 and W911NF-12-1-0607. The views and conclusions contained in this document are those of the authors and should not be interpreted as representing the official policies, either expressed or implied, of the ARO or the US Government. The US Government is authorized to reproduce and distribute reprints for government purposes notwithstanding any copyright notation herein. M.R. acknowledges support from the Netherlands Organization of Scientific Research (NWO) under Veni grant VI.Veni.212.223.

\section*{Additional information}
\paragraph*{Data availability} The raw data and analysis that support the findings of this study are available at the Zenodo repository: https://doi.org/10.5281/zenodo.7070819.

\bibliography{literature}
\bibliographystyle{apsrev4-1}

\newpage
\newpage
\clearpage

\appendix
\section{Setup}
\label{Appendix A: setup}
\noindent The measurements were carried out in an Oxford Triton dry dilution refrigerator with a base temperature around $10$~mK. The DC part of the setup consists of two in-house built, battery-powered SPI racks, containing digital-to-analog converters. The voltage pulses applied to the sample are generated by two Tektronix AWG5014 arbitrary waveform generators and sent to the sample via coaxial lines, connected via a bias tee on the printed circuit board with a cut-off frequency of $3$~Hz. The response of the left charge sensor was monitored with an RF reflectometry setup at a resonance frequency of $f=214$~MHz, containing an in-house fabricated NbTiN inductor. The reflectometry signal is amplified at the $4$-Kelvin plate and demodulated using an in-house built SPI demodulation rack. The sensor response of the right charge sensor was converted to voltage through a home-built baseband amplifier. This latter signal was not used for single-shot readout.

\section{Data analysis}
\label{Appendix B: data analysis}
\noindent We smoothen the data by means of boxcar averaging; we average each data point with the $25$ points around it. Subsequently, we assign each data trace `$0$', or `$1$' by means of relative thresholding (adjusting the threshold based on the signal of a readout segment during which the first dot is always occupied with one electron). During the measurements, we encountered a timing problem between the AWG and the digitizer, which resulted in an offset of up to $50$~\textmu s in the readout traces for different shuttle rounds, see Fig.~\ref{fig: Supp_traces}. To ensure a similar readout duration for each data trace, we cut off the start and end of each data trace by the same number of points.

\section{Simulation of spin-flip errors from shuttling}
\label{Appendix C: simulations}
\noindent We aim to bound the spin-flip probability per hop by comparing the experimental data with numerical simulations. In our simulations, we consider spin relaxation from spin up to spin down and a finite spin-flip probability during shuttling, which we assume to be identical for each hopping event. Since the time between events is expected to be on the order of $T_2^*$ or larger, we neglect the phase evolution in our simulations and restrict ourselves to the dynamics of the spin up and spin down probabilities $P_{\ket{\uparrow}}$, $P_{\ket{\downarrow}}$. In this model, relaxation for quantum dot $j$ reads in the basis $\{\ket{\uparrow}$,$\ket{\downarrow}\}$

\begin{equation}
R_j = \begin{pmatrix}
\exp[-t_{\mathrm{wait}}/T_{1,j}]  & 0\\
1-\exp[-t_{\mathrm{wait}}/T_{1,j}]     & 1
\end{pmatrix}.
\end{equation}

The spin-flip probability matrix, with spin-flip probability $\beta$ is:

\begin{equation}
F = \begin{pmatrix}
1-\beta & \beta\\
\beta   & 1-\beta
\end{pmatrix}.
\end{equation}

The time spent in each quantum dot depends on the total time in the array, $t_{\textrm{array}}$ and the number of shuttle rounds $n$ as $t_{\mathrm{wait}} = t_{\textrm{array}}/(6n)$, where the factor $6$ accounts for the six stages of the shuttling sequence (dot 1, dot 2, dot 3, dot 3, dot 2, dot 1). After $n$ shuttle rounds, the final spin probabilities are

\begin{equation}
\begin{pmatrix}
P_{\ket{\uparrow},\textrm{final}}\\
P_{\ket{\downarrow},\textrm{final}}
\end{pmatrix} = (R_1 F R_2 F R_3 R_3 F R_2 F R_1)^n 
\begin{pmatrix}
P_{\ket{\uparrow},\textrm{init}}\\
P_{\ket{\downarrow},\textrm{init}}
\end{pmatrix}.
\end{equation}

\begin{figure}[t]
    \centering
    \includegraphics{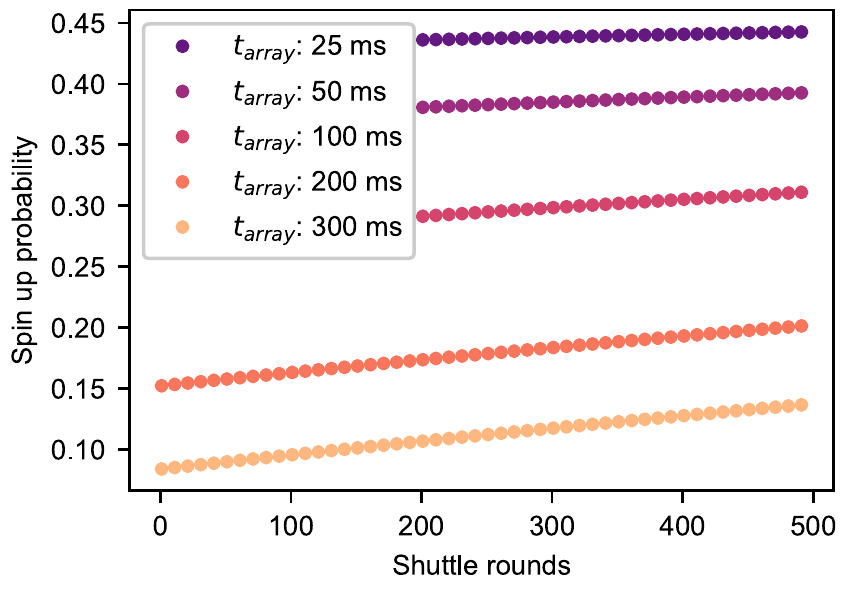}
    \caption{Simulation of spin-up probability at the end of a shuttling sequence through three dots, considering a $0.01\%$ probability of a spin flip per hop and $T_1$ decay during the wait times in the dots, for different $t_{\mathrm{array}}$.}
    \label{fig: Supp_flip_chance}
\end{figure}

\begin{figure*}[t]
    \centering
    \includegraphics[width=\textwidth]{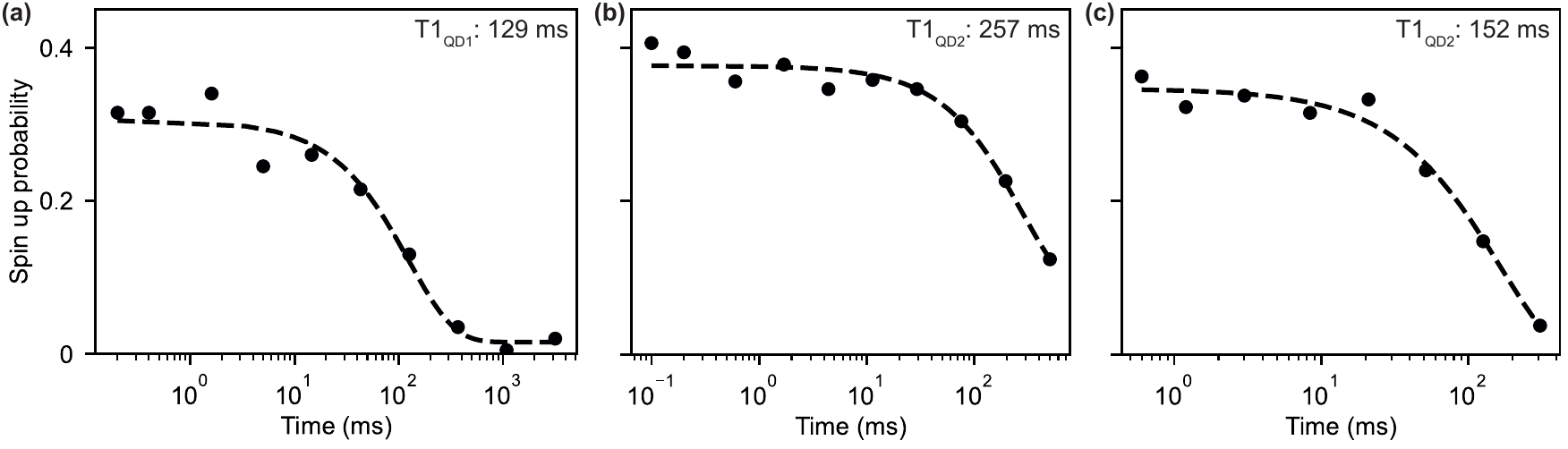}
    \caption{Relaxation times for the separate quantum dots. Spin-up probability of a randomly loaded spin as a function of waiting time in (a) dot 1, (b) dot 2 and (c) dot 3 at a magnetic field of $1.3$~T. Each data point is averaged $300$, $500$ and $400$ times for dot $1$, $2$ and $3$ respectively. Fitting the data to an exponential decay yields relaxation times of $T_{1,1}=129\pm33$~ms, $T_{1,2}=257\pm79$~ms and $T_{1,3}=152\pm48$~ms. The relatively large error bars are a result of the sparsity of data points for longer waiting times.}
    \label{fig: Supp_T1_separate_dots}
\end{figure*}

Fig.~\ref{fig: Supp_flip_chance} shows numerical simulation results considering a spin-flip probability per hop of $0.01\%$, spin relaxation during the wait times between hops, with time constants $T_{1,1}$, $T_{1,2}$ and $T_{1,3}$, as measured in Fig.~\ref{fig: Supp_T1_separate_dots} and initial probabilities $P_{\ket{\uparrow},\textrm{init}}=P_{\ket{\uparrow},\textrm{init}}=0.5$. The simulation shows the resulting spin-up probability as a function of the number of shuttle rounds $n$. Especially for the longer array times, the spin up probability follows an upward trend with the number of shuttle rounds. In contrast, no upward trend can be seen in the experiment of Fig.~\ref{fig: Figure_3}, indicating that the experimental spin-flip probability per hop is below $0.01\%$.

\section{Shuttling for 500 rounds}
\label{Appendix D: shuttle 500 rounds}
\begin{figure}[b]
    \centering
    \includegraphics{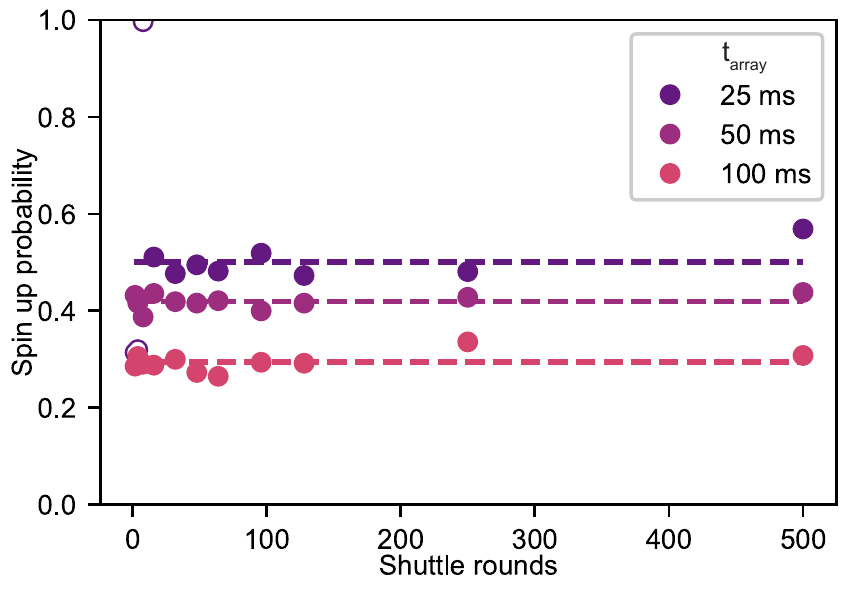}
    \caption{Shuttling for $500$ rounds through an array of three quantum dots ($2000$ hops). Each data point is the average of $1000$ single-shot traces, taken for three different total times in the quantum dot array: $25$, $50$ and $100$~ms. As there was a shift in the readout position during the data collection for the $25$~ms case, the first three data points have a calibration error. These outliers are plotted as open circles, yet are not included in the data analysis (see also Fig.~\ref{fig: Supp_traces}).}
    \label{fig: Supp_500_shuttles}
\end{figure}

\begin{figure*}[t]
    \centering
    \includegraphics{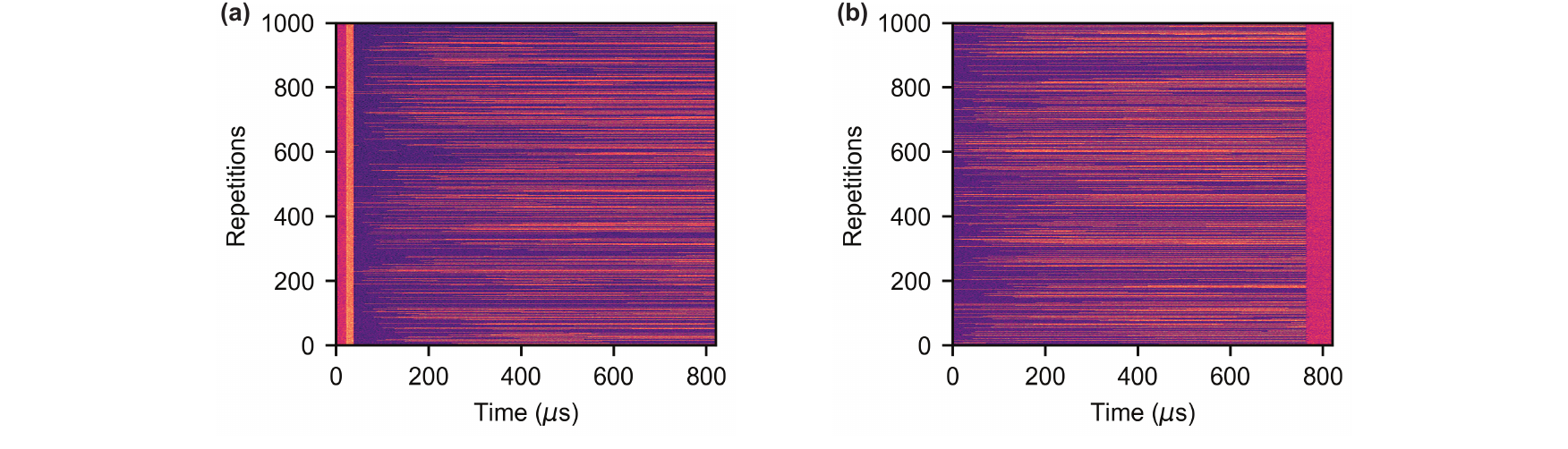}
    \caption{Single-shot traces for three different data points in Fig.~\ref{fig: Supp_500_shuttles}. Single-shot traces showing the sensing dot response of the readout segment after shuttling the electron (a) $250$ and (b) $500$ rounds through the array respectively for $t_{\mathrm{array}}=50$~ms. (a) The digitizer timing is slightly offset with respect to the readout pulse; the last two shuttling steps are still visible on the left side of each trace. (b) The digitizer timing is offset with respect to the readout pulse; the start of the compensation pulse is seen on the right side of each trace.}
    \label{fig: Supp_traces}
\end{figure*}

\noindent Data for shuttling through three quantum dots for up to $500$ shuttling rounds is shown in Fig.~\ref{fig: Supp_500_shuttles}.\\

\section{Relaxation times in each quantum dot}
\label{Appendix E: relaxation times}
\noindent Fig.~\ref{fig: Supp_T1_separate_dots} shows data from which the spin relaxation times are extracted for each dot separately.

\section{Charge shuttling over many rounds}
\label{Appendix F: shuttle many rounds}
\begin{figure}[h!]
    \centering
    \includegraphics{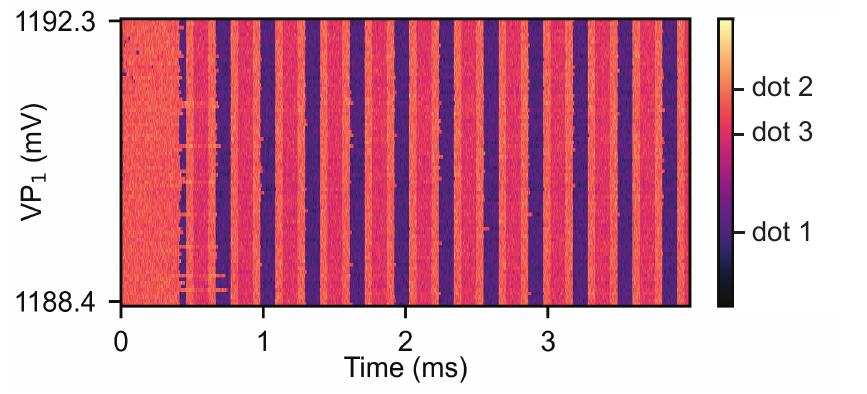}
    \caption{Charge shuttling data showing the $\approx 11$ first shuttling rounds of a  shuttling sequence of 78 rounds, showing the sensor response for 100 repetitions (each trace has an offset with respect to the previous trace of 0.04 mV in VP1). The time between hops is about $100$~\textmu s for dots 1 and 3 ($50$~\textmu s for dot 2), roughly four times shorter compared to the data in Fig.~\ref{fig: Figure_2}. By zooming in on the first 11 shuttling rounds, latching effects during loading from the reservoir can be seen in a subset of the traces. Even during the shuttling rounds, it can be seen that in some cases the electron passes from one dot to the next with a small delay. This is not expected to impact the spin polarization but may impact the spin phase, in case the spin splittings vary between dots.}
    \label{fig: Supp_many_shuttles}
\end{figure}

\noindent Fig.~\ref{fig: Supp_many_shuttles} shows charge sensing data where a charge is shuttled through the array of three quantum dots many times, four times faster than in Fig.~\ref{fig: Figure_2}.

\begin{figure*}[b]
    \centering
    \includegraphics{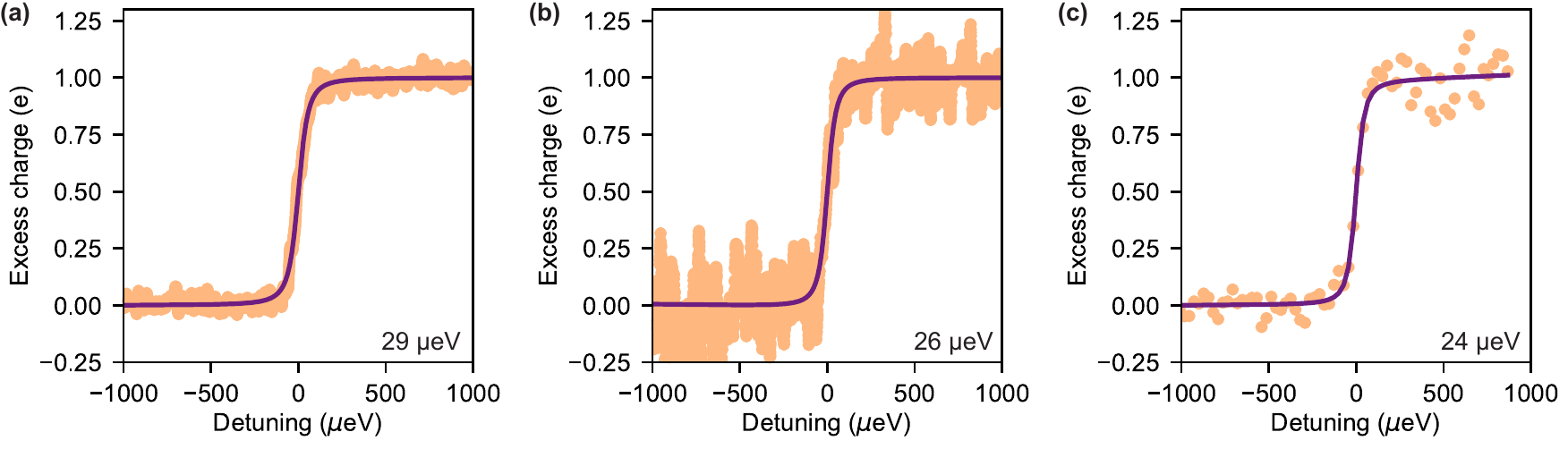}
    \caption{Tunnel coupling. Scan along the interdot detuning axes of (a) QD1-QD2, (b) QD2-QD3 and (c) QD3-QD4 (right sensing dot).}
    \label{fig: Supp_tunnel_coupling}
\end{figure*}

\section{Tunnel coupling}
\label{Appendix G: tunnel coupling}
Measurements of the tunnel coupling between the subsequent quantum dots are shown in Fig.~\ref{fig: Supp_tunnel_coupling}. The data has been fit with the model from~\cite{dicarlo_differential_2004, van_diepen_automated_2018}. The tunnel coupling extracted from the fit for each interdot transition is given in \textmu eV. The lever arm of $P5$ to dot 5 was extracted from Coulomb diamonds. The lever arms of the virtual plunger gates to the respective dot potentials was successively determined from charge stability diagrams (by comparing the relative capacitive coupling of each virtual gate to that of gates with previously extracted lever arms). For the fit, we used an electron temperature of $75$~mK, which is similar to what was measured in previous and later cooldowns. The signal-to-noise ratio for the QD2-QD3 interdot transition is relatively low, because of the distance to the sensing dot and the sensing-dot sensitivity.

\clearpage
\end{document}